# Clustering of Mobile Ad Hoc Networks: An Adaptive Broadcast Period Approach

Damianos Gavalas[1], Grammati Pantziou[2], Charalampos Konstantopoulos[3], Basilis Mamalis[2]

[1] Department of Cultural Technology and Communication, University of the Aegean, Greece, dgavalas@aegean.gr

[2] Department of Informatics, Technological Education Institute of Athens, Greece, {pantziou, vmamalis}@teiath.gr

[3] Computer Technology Institute, Patras, Greece, konstant@cti.gr

**Abstract** - Organization, scalability and routing have been identified as key problems hindering viability and commercial success of mobile ad hoc networks. Clustering of mobile nodes among separate domains has been proposed as an efficient approach to address those issues. In this work, we introduce an efficient distributed clustering algorithm that uses both location and energy metrics for cluster formation. Our proposed solution mainly addresses cluster stability, manageability and energy efficiency issues. Also, unlike existing active clustering methods, our algorithm relieves the network from the unnecessary burden of control messages broadcasting, especially for relatively static network topologies. This is achieved through adapting broadcast period according to mobile nodes mobility pattern. The efficiency, scalability and competence of our algorithm against alternative approaches have been demonstrated through simulation results.

## I. INTRODUCTION

Wireless communication and the lack of centralized administration pose numerous challenges in mobile wireless ad-hoc networks (MANETs) [6]. Node mobility results in frequent failure and activation of links, causing a routing algorithm reaction to topology changes and hence increasing network control traffic [2]. Ensuring effective routing and QoS support while considering the relevant bandwidth and power constraints remains a great challenge. Given that MANETs may comprise a large number of MNs, a hierarchical structure will scale better [5].

Hence, one promising approach to address routing problems in MANET environments is to build hierarchies among the nodes, such that the network topology can be abstracted. This process is commonly referred to as *clustering* and the substructures that are collapsed in higher levels are called *clusters* [12]. The concept of clustering in MANETs is not new; many algorithms that consider different metrics and focus on diverse objectives have been proposed [12]. However, most existing algorithms fail to guarantee stable cluster formations. More importantly, they are based on periodic broadcasting of control messages resulting in increased consumption of network traffic and mobile hosts (MH) energy.

In this article, we introduce a distributed algorithm for efficient and scalable clustering of MANETs that corrects the two aforementioned weaknesses. The main contributions of the algorithm are: fast completion of clustering procedure, where both location and battery power metrics are taken into account; derived clusters are sufficiently stable, while cluster scale is effectively controlled so as not to grow beyond certain limits; minimization of control traffic volume, especially in relatively static MANET environments.

The remainder of the paper is organized as follows: Section II provides an overview of clustering concepts and algorithms. Section III describes the details of our Adaptive Broadcast Period algorithm and Section IV discusses simulation results. Finally, Section V concludes the paper and draws directions for future work.

## II. CLUSTERING

In clustering procedure, a representative of each subdomain (cluster) is 'elected' as a *cluster head* (CH) and a node which serves as intermediate for inter-cluster communication is called *gateway*. Remaining members are called *ordinary nodes*. The boundaries of a cluster are defined by the transmission area of its CH. With an underlying cluster structure, non-ordinary nodes play the role of dominant forwarding nodes, as shown in Figure 1.

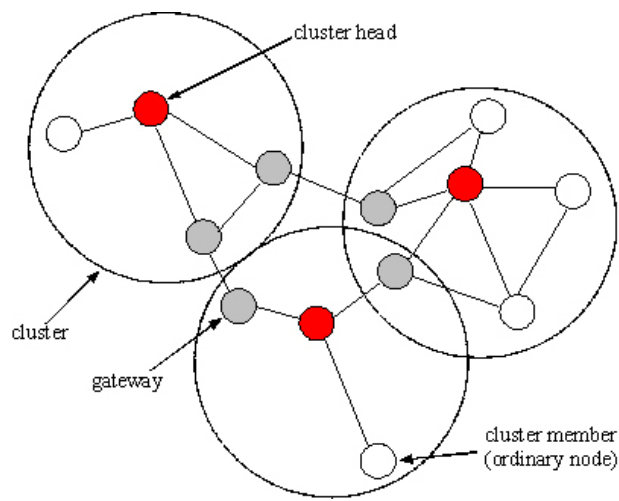

Figure 1. Cluster heads, gateways and ordinary nodes in mobile ad hoc network clustering.

Cluster architectures do not necessarily include a CH in every cluster. CHs hold routing and topology information, relaxing ordinary MHs from such requirement; however, they represent network bottleneck points. In clusters without CHs, every MH has to store and exchange more topology information, yet, that eliminates the bottleneck of CHs. Yi et al. identified two approaches for cluster formation, *active* clustering and *passive* clustering [10]. In active clustering, MHs cooperate to elect CHs by periodically exchanging information, regardless of data transmission. On the other hand, passive clustering suspends clustering procedure until data traffic commences [11]. It exploits on-going traffic to propagate "cluster-related information" (e.g., the state of a node in a cluster, the IP address of the node) and collects neighbor information through promiscuous packet receptions.

Passive clustering eliminates major control overhead of active clustering, still, it implies larger setup latency which might be important for time critical applications; this latency is experienced whenever data traffic exchange commences. On the other hand, in active clustering scheme, the MANET is flooded by control messages, even while data traffic is not exchanged thereby consuming valuable bandwidth and battery power resources.

Recently multipoint relays (MPRs) have been proposed to reduce the number of gateways in active clustering. MPR hosts are selected to forward broadcast messages during the flooding process [7]. This technique substantially reduces the message overhead as compared to a typical flooding mechanism, where every node retransmits a message when it receives its first copy. Using MPRs, the Optimized Link State Routing (OLSR) protocol can provide optimal routes, and at the same time minimize the volume of signaling traffic in the network [1]. An efficient clustering method should be able to partition a MANET quickly with little control overhead. Due to the dynamic nature of MANETs, optimal cluster formations are not easy to build. To this end, two distributed clustering algorithms have been proposed: Lowest ID algorithm (LID) [10] and Highest Degree algorithm (HD) [10]. Both of them belong to active clustering scheme.

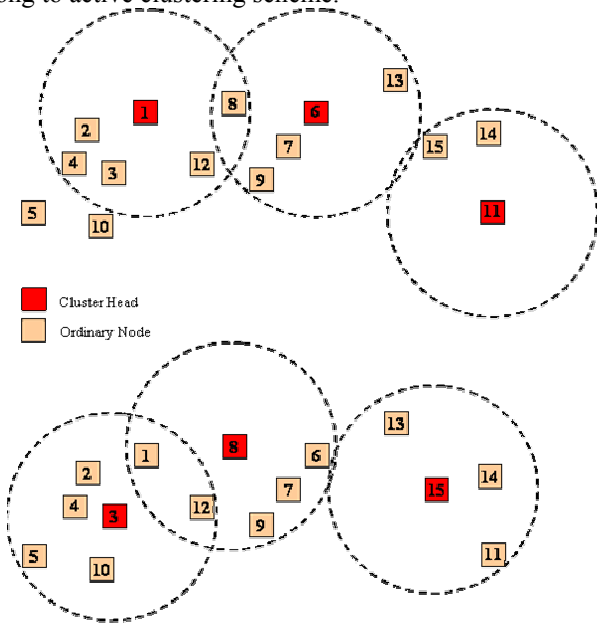

Figure 2. LID vs. HD algorithms clustering.

In LID algorithm, each node is assigned a unique ID. Periodically, nodes broadcast the list of nodes located within their transmission range (including themselves) through a 'Hello' control message. The lowest-ID node in a neighborhood is then elected as the CH; nodes which can 'hear' two or more CHs become gateways, while remaining MHs are considered as ordinary nodes. In HD algorithm, the highest *degree* node in a neighborhood, i.e. the node with the largest number of neighbors is elected as CH. Figure 2 compares LID vs. HD algorithm approaches.

LID method is a quick clustering method, as it only takes two 'Hello' message periods to decide upon cluster structure and also provides a more stable cluster formation than HD. In contrast, HD needs three 'Hello' message periods to establish a clustered architecture [3]. In HD method, losing contact of a single node (due to MH movement), may cause failure of the current CH to be re-elected. On the other hand, HD method can get fewer clusters than LID, which is more advantageous in large-scale network environments.

In current clustering schemes, stability and cluster size are very important parameters; however, reducing the number of clusters does not necessarily result in more efficient architectures. A CH may end up dominating so many MHs that its computational, bandwidth and battery resources will rapidly exhaust. Therefore, effective control of cluster size is another crucial factor.

Summarizing, both LID and HD algorithms use exclusively location information to form clusters and elect CHs. In a more recent approach, Li et al proposed Vote-based Clustering (VC) algorithm, where CH elections are based not purely on location but also on the battery power level of MHs [3]. In particular, MHs with high degree (large number of neighbors) and sufficient battery power are elected as CHs. However, simulations have shown that the combination of position and power information in clustering procedure results in frequent CH changes, i.e. overall cluster structure instability [3]. In a MANET that uses cluster-based services, network performance metrics such as throughput, delay and effective management are tightly coupled with the frequency of cluster reorganization. Therefore, stable cluster formation is essential for better management and QoS support.

In addition, LID, HD and VC algorithms share a common design characteristic which derives from their active clustering origin. Cluster formation is based on the periodic broadcast of 'Hello' signaling messages. In cases where MHs are relatively static (e.g. in collaborative computing, on-the-fly conferencing, etc), periodic 'storms' of control messages only occur to confirm that cluster structure established in previous periods should remain unchanged. These unnecessary message broadcasts not only consume network bandwidth, but valuable battery power as well.

### III. ADAPTIVE BROADCAST PERIOD (ABP) ALGORITHM

Our Adaptive Broadcast Period (ABP) algorithm aspires to correct the inefficiencies of existing active clustering algorithms (LID, HD and VC). Emphasis is given on three directions:

- A quick method for cluster formation is needed; required speed though should not be achieved at the expense of instable cluster configurations. To meet this objective, we modify VC algorithm so as to avoid frequent CH 're-elections'.
- Cluster sizes should be controlled so as not to derive too large neither too small clusters.
- Control messages broadcast period should be dynamically adapted to avoid unnecessary message exchanges when the mobility pattern of MHs is such that network topology is relatively static.

The methodology chosen to achieve the three aforementioned objectives is detailed in the following sections.

*A. Cluster Formation*

Similarly to VC and unlike LID and HD protocols, both position and battery power metrics are considered in CH election. However, emphasis has been given to prevent frequent CH changes and prolong the average lifetime of CH serving time and cluster membership, therefore, meeting the requirement for steadier cluster formations.

**Network Model**

A MANET can be divided into several overlapped clusters. A cluster comprises of a subset of nodes that communicate via their assigned CH. The network is modeled as an undirected graph $G$ ($V,E$) where V denotes the set of all MHs (*vertices*) in the MANET and $E$ denotes the set of links or *edges* ($i, j$) where $i, j \in V$. Each link signifies that two MHs are within the transmission range of each other. Let $S_i$ be the set of MHs that can be reached by node $i$. We assume every link is bi-directional so that link ($i, j$) exists if and only if $j \in S_i$.

Each MH has a unique identifier (MH_ID), which is a positive integer. MHs also hold information about the identity of their assigned CH (CH_ID). CHs are easily identified by their identical MH_ID and CH_ID values.

Control information is communicated through 'Hello' messages, transmitted on the common wireless channel. Every MH acquires information from incoming 'Hello' message sent by its neighbors. We assume that only when two MHs lie within mutual transmission range, they can communicate directly with each other, i.e. a bi-directional link exists. Another attribute of MHs is their battery power level (percentage of remaining over full battery power), which is a positive integer, $0 \leq b \leq 100$. We assume linear decrease of $b$ over time; naturally, battery energy of CHs exhausts faster than ordinary MHs as they serve a number of MHs, forwarding messages on their behalf.

**Clustering Algorithm**

Our clustering algorithm considers both location and power information to partition a MANET into separate clusters. In this context, we introduce the concept of "cluster head competence" (CHC) which represents the competence of a MH to undertake the role of a CH.

The format of a typical 'Hello' message is shown in Figure 3. Each 'Hello' message includes identifications of its sender (MH_ID) and sender's assigned CH (CH_ID). CCH represents a weighted sum of sender's degree (number of neighbors) and its battery power level. Finally, the 'Option' message field is used for cluster size management purposes (see subsection B), and Broadcast Period (BP) field is used to adapt the broadcast period within a particular cluster (see subsection C).

| MH_ID | CH_ID | CHC | Option | BP |
|---|---|---|---|---|
| 8 bit | 8 bit | 8 bit | 4 bit | 8 bit |

Figure 3. 'Hello' packet format.

CHC values are calculated according to the following equation:

$$CHC = (c_1 \times d + c_2 \times b) - p \qquad (1)$$

- $c_1$, $c_2$: weighted coefficients of MH degree and battery availability, respectively ($0 \leq c_1, c_2 \leq 1, c_1 + c_2 = 1$);
- $d$: Number of neighbors (degree of MH);
- $b$: Remaining battery lifetime (percentage of remaining over full battery power);
- $p$: 'handover' penalty coefficient (explained in the following subsection).

The algorithm's execution involves the following steps:
(1) Each MH sends a 'Hello' message randomly during a 'Hello' cycle. If a MH has just joined the MANET, it sets CH_ID value equal to a negative number. That signifies a MH is not a member of any cluster and has no knowledge of whether it is within transmission radius of another MH.
(2) Each MH counts how many 'Hello' messages it received during a 'Hello' period, and considers that number as its own degree ($d$).
(3) Each MH broadcasts another 'Hello' message, setting CHC field equal to the value calculated from Equation (1).
(4) Recording received 'Hello' messages during two 'Hello' cycles, each MH identifies the sender with highest CHC value and thereafter considers it as its CH.

In the next 'Hello' cycle, CH_ID value will be set to elected CH's ID value. In the case of two or more MHs having the same lowest CHC value, the one with the lowest ID is 'elected' as CH. Following the aforementioned algorithm steps, clustering procedure is completed within two 'Hello' cycles.

ABP execution steps are illustrated in Figure 4. Table I presents how CHC values are calculated, where the coefficients of equation (1) are set to $c_1 = 0.4$ and $c_2 = 0.6$.

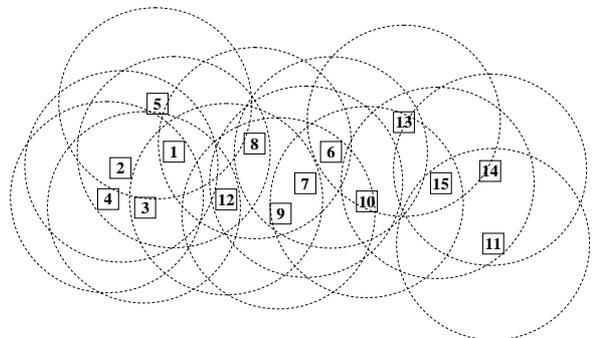

(a)

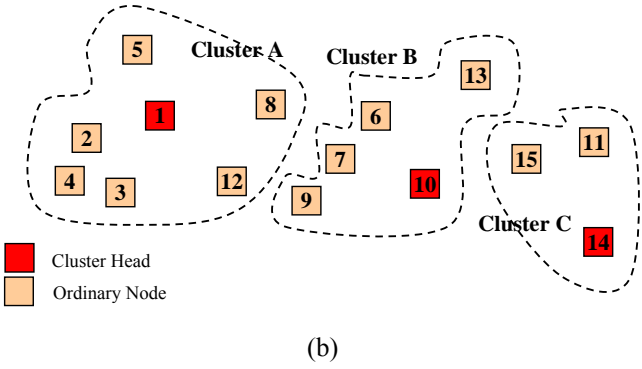

(b)

Figure 4. Illustration of ABP execution: (a) Original placement of mobile nodes on the plane (dashed circles indicate nodes transmission range), (b) Cluster formation based on ABP clustering (CHC values calculation is based on the figures provided in Table I).

TABLE I. CALCULATION OF *CHC* VALUES IN ABP (WHERE $c_1$ = 0.4, $c_2$ = 0.6 AND $p$ = 1)

| MH ID | d | b | CHC |
|---|---|---|---|
| 1 | 6 | 4 | 3,8 |
| 2 | 4 | 5 | 3,6 |
| 3 | 4 | 3 | 2,4 |
| 4 | 3 | 4 | 2,6 |
| 5 | 2 | 2 | 1 |
| 6 | 5 | 4 | 3,4 |
| 7 | 5 | 2 | 2,2 |
| 8 | 5 | 1 | 1,6 |
| 9 | 5 | 4 | 3,4 |
| 10 | 5 | 5 | 4 |
| 11 | 2 | 4 | 2,2 |
| 12 | 5 | 2 | 2,2 |
| 13 | 3 | 4 | 2,6 |
| 14 | 2 | 7 | 4 |
| 15 | 4 | 2 | 1,8 |

**Securing Cluster Stability**

Notably, our clustering algorithm extends the ideas of HD approach so as to include the battery energy metric in CH election process (when setting $c_1$ = 1, $c_2$ = 0 and $p$ = 0 in Equation (1), the two algorithms coincide). This similarity implies that our proposed algorithm runs the risk of providing instable cluster formations (cluster instability has been identified as the main weakness of HD algorithm [3]), i.e. clusters are sensitive to hosts mobility.

According to the preceding description of our algorithm steps, CH re-election occurs when an ordinary MH claims higher CHC value compared to the current CH. For instance, that is likely to happen when: (a) a cluster member relocates away of its cluster perimeter, (b) a new MH moves within a cluster boundary, (c) the current CH presents slightly lower power level than an ordinary MH. Such re-election could trigger a global cluster reconfiguration process and massive transfers of routing data among elected CHs.

To correct this inefficiency, we introduce a penalty coefficient *p* in the calculation of CHC value, as shown in Equation (1). The value of *p* is set to an integer value ($p > 0$) for ordinary MHs, while *p* = 0 for CHs. Assigning an appropriate value to *p*, we prevent MHs with slightly higher degree or lower battery power to that of current CHs to take up the role of CH, thereby avoiding unnecessary handovers. In other words, CH re-elections occur only in the event of major modifications of MANET topology structure (e.g. current CHs' degree has significantly decreased) or in cases where CHs future engagement on packet forwarding activity will soon cause their battery exhaustion.

*B. Cluster Size Management*

The objective of clustering algorithms is to partition the network into several clusters. Optimal cluster size is dictated by the tradeoff between spatial reuse of the channel (which drives towards small clusters) and delay minimization (which drives towards large clusters) [5]. In addition, large clusters lead to rapid exhaustion of CH battery power, while CHs represent network bottleneck points. On the other hand, small cluster sizes lead to formation of multiple clusters, implying growth of routing information and also network topology which is difficult to manage.

To address the issue of efficient cluster size management, we propose an adaptive cluster load balance method. The 'Option' field of 'Hello' packet (see Figure 3) is used for that purpose. CHs set the value of 'Option' field equal to the number of their dominated MHs. In contrast, ordinary MHs reset the 'Option' field value to 0. The number of MHs dominated by a single CH is not allowed to exceed a specified threshold *T*. The value of 'Option' field is of importance for MHs currently not belonging to any cluster or not being dominated by the CH that issued the 'Hello' message. In such cases, if 'Option' value equals *T*, MHs cannot request membership to the CH that broadcasted the 'Hello' message. As a result, potential CH bottlenecks are prevented and cluster sizes are moderated.

In addition, the above described cluster size management method guarantees balanced load among various clusters. Resource consumption and data traffic is fairly distributed among network clusters, and does not burden certain clusters against others.

*C. Dynamically Adaptive Control Messages Broadcast Period*

A principal consideration of our Adaptive Broadcast Period (ABP) algorithm is to reduce the number of control messages circulated within the MANET. Minimization of message broadcasts ensures bandwidth savings and conserves computational resources and battery power not only on elected CHs but on ordinary nodes also.

The idea behind controlling the volume of broadcast messages is based on the realistic hypothesis that ad-hoc networks are not always highly mobile. This is usually the case in MANETs facilitating communication of mobile users in convention centers, conferences or electronic classrooms. Existing active clustering algorithms involve periodic broadcast of 'Hello' messages to sense potential topological differences between two successive 'Hello' periods. When considering relatively static MANET topologies though, such modifications seldom occur. Namely, bandwidth and power resources are consumed only to verify that existing clustering configurations are still valid.

ABP algorithm corrects this clear inefficiency by dynamically adjusting 'Hello' broadcast period (BP). In particular, BP duration depends on the current mobility pattern of MHs. For highly mobile MHs, BP is shortened, i.e. message broadcasts are frequent enough to maintain consistent and accurate topology information. However, when mobility rate (MR) is low (i.e., MHs position on the plane does not considerably change over time relatively to their neighbors position), BP is lengthened, relaxing the MANET from unnecessary control message storms.

Yet, it is essential to guarantee that all individual cluster members share the same BP value. Should permission to request adaptation of BP is granted to all MHs, that will soon lead to serious BP synchronization problem: MHs are likely to receive simultaneous BP adaptation requests from different MHs. As a result, members of the same cluster will adjust their BP to different time-spans, which will severely affect the validity of CH 'election' process described above. Hence, in ABP algorithm, only CHs are entitled to issue BP adaptation requests to their dominated MHs. In case of node migration to a neighboring cluster, its new CH informs the node about the BP of the local cluster.

Most existing methods for estimating nodes mobility rate pose the requirement for GPS card with sufficient accuracy mounted on every mobile node. We propose an alternative method for measuring MR which relaxes mobile nodes from such requirement.

Each CH $v$ measures its neighborhood MR through contrasting the topology information it obtains during successive BPs. CHs maintain a short 'topology history table' (THT); THT rows comprise vectors representing the IDs of neighboring nodes, where each THT row refers to different BP. Calculated MR value actually represents the mean 'vector distance' among vectors recorded by $v$ during the latest $n$ BPs (where $n$ is a small integer in order to minimize memory requirement):
$$MR_t = (1/n)\sum_{i=0}^{n-1}\left|\overline{THT_{t-iBP} - THT_{t-(i+1)BP}}\right|,$$
where $t$ denotes the current time.

Figure 5 illustrates how mobile node with ID = 1 moves on the plane; as a result of that movement (and the movement of other network nodes), its neighboring nodes (i.e. those within its transmission range) differ at the end of every BP. For this particular example, the 'neighborhood vectors' of node #1 at the end of four successive BPs: are $THT_1 = \{2, 3, 4, 5, 8, 12\}$, $THT_2 = \{2, 3, 5, 9, 12\}$, $THT_3 = \{2, 3, 5\}$, $THT_4 = \{3, 8, 12, 14\}$. Hence, the mobility rate of node #1 within this period of time is given by:

$M_1 = (\left|\overline{THT_4 - THT_3}\right| + \left|\overline{THT_3 - THT_2}\right| + \left|\overline{THT_2 - THT_1}\right|)/3 =$
$(5+2+3)/3 = 3.33.$

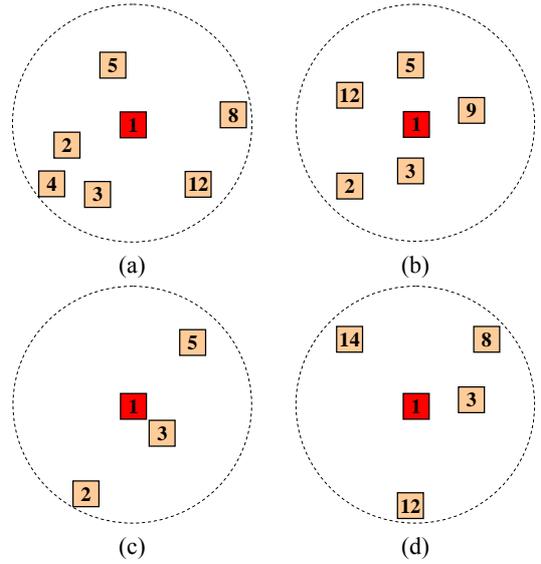

Figure 5. Neighboring nodes of node with ID = 1 during four successive BPs.

A main objective of ABP algorithm is to minimize control traffic overhead during clustering maintenance phase, which highly depends on BP duration (i.e. frequency of broadcasting 'Hello' control packets). To achieve that, CHs measure the mean mobility rate of their attached cluster members $MR_c$ (following the above described method) and accordingly adapt the 'Hello' broadcast period BP within their cluster. It is also guaranteed that BP duration always lies between two boundaries: $BP_{min} \leq BP \leq BP_{max}$; at startup, BP is globally set to $BP_{min}$.

IV. PERFORMANCE SIMULATION AND ANALYSIS

Our simulation work attempts to compare the performance of ABP against LID, HD and VC algorithms in terms of signaling traffic, cluster stability and variance of MHs energy level. Simulations have been performed using the NS-2 simulation package [9].

A square terrain of 600m × 600m is assumed. The number of MHs moving within the square space varies from 20 to 120. At startup, MHs are randomly positioned on the plane. MHs move with speed 0 - 15m/s, on random direction. At the event of reaching the terrain boundary, MHs are bounced back. The BP duration is set to 5ms for LID, HD and VC approaches while for ABP algorithm it is dynamically adjusted according to MHs mobility behavior. Initial remaining battery time of MHs is randomly set between 20 and 100 units; energy is assumed to be linearly decreased for ordinary nodes, while for CHs it depends on the number of their attached cluster members. Each simulation run lasts 3 minutes; simulation results presented below have been averaged over 5 runs. Regarding ABP algorithm's execution parameters, CHC values are calculated for $c_1 = c_2 = 0.5$, while the value of penalty coefficient is set to $p = 2$. The maximum number of nodes that may be dominated by a single CH is set to $T = 10$. CHs measure mobility rate through contrasting the topology information they obtain during $n = 5$ successive BPs.

Figure 6 illustrates the average number of control messages exchanged over the simulation runs. In LID, HD and VC algorithms, 'Hello' messages are periodically broadcasted, hence, their performance results coincide. As expected though, ABP clearly outperforms the three alternative approaches, especially when MHs exhibit low mobility.

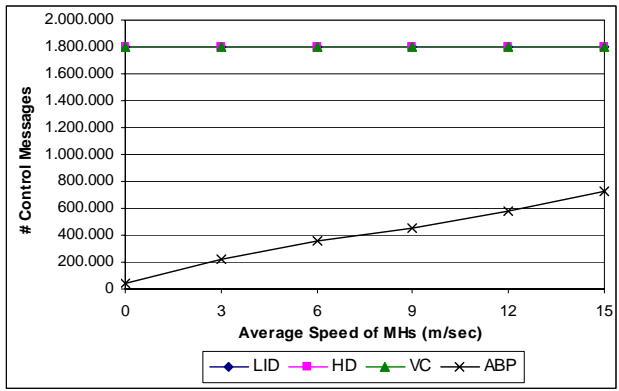

Figure 6. Average number of control messages exchanged (for 50 MHs).

However, control messages ('Hello' packets) are not of equal size in all four examined approaches. In particular, 'Hello' packet sizes are 8, 8, 32 and 36 bits for LID, HD, VC and ABP respectively. That certainly affects the scalability of the clustering algorithms, as shown in Figure 7. Thus, in terms of the overall control traffic overhead, ABP is shown to perform better than LID and HD when the average speed of MHs is not larger than 7 m/sec, while it presents clearly better results than VC. Here again, the performance of LID and HD coincide, due to their identical BP duration and control packet size.

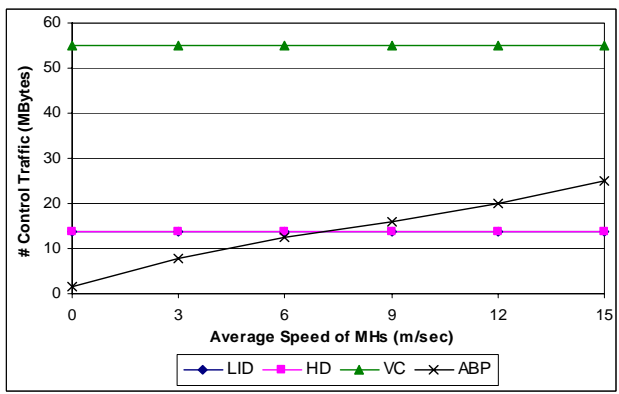

Figure 7. Control traffic volume (for 50 MHs)

Figure 8 compares the average number of CH changes, which is an indicator of the overall cluster structure stability (the more frequent the CH changes, the less stable clusters are). As expected, LID performs better than HD as the former exclusively uses ID and the latter node degree information to decide upon cluster structure. VC performs even worse than HD as the inclusion of power level metric in CHs election dictates that CHs with insufficient power level give up their CH role. However, ABP is marginally outperformed by LID on account of the penalty coefficient which prevents frequent CH re-elections.

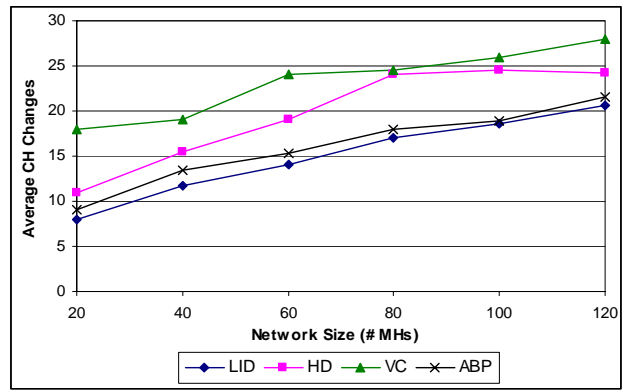

Figure 8. Average number of CH changes (for average speed of 5 m/sec).

Finally, Figure 9 illustrates the variance of power level among MANET's MHs. Large variance values indicate that specific nodes are engaged on CH role for long periods, hence, their energy level soon falls far below the average. This simulation result highlights the main limitation of LID algorithm: in LID, CHs election is biased in favor of nodes with low ID values; these nodes are likely to serve as CHs for long time and their energy supply rapidly depletes. ABP results in a more fair distribution of energy consumption compared to LID and HD as it takes into account remaining power level for CHs election. However, ABP demonstrates marginally worse performance than VC, as the inclusion of penalty coefficient extends CHs serving time, i.e. it prevents CHs with slightly lower battery power to give up their CH role. That represents an interesting trade-off between stable and energy-balanced clustering.

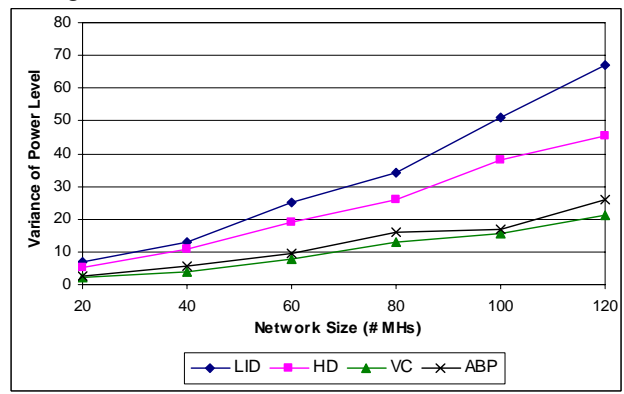

Figure 9. Variance of energy level among MHs (for average speed of 5 m/sec).

## V. CONCLUSIONS – FUTURE WORK

In this article, we introduced a novel active clustering algorithm. Its contributions, compared to existing solutions, are summarized in the following: (a) clustering procedure is completed within two 'Hello' cycles; (b) both location and battery power metrics are taken into account in clustering process; (c) derived cluster formations exhibit enhanced stability by preventing unnecessary CH re-elections; (d) cluster sizes are controlled so as not to expand beyond a specified threshold; (e) for relatively static network topologies, control traffic volume is minimized; (f) fast packet forwarding and delivery is enabled, as clusters are pro-actively formed and topology information is available when actual user data exchange is required. The abovementioned contributions are achieved at the expense of slightly increased control packet sizes which may result in increased control traffic volume in highly mobile environments.

Simulation results demonstrated that APB algorithm achieves cost-effective clustering in terms of signaling traffic, especially for MANETs with low to moderate mobility rate. Also, it represents a balanced solution between cluster stability and energy efficiency compared to existing approaches.

As a future extension, we intend to incorporate mobility metric in the calculation of cluster head competence, and also introduce a mobility prediction method (e.g. similar to [8]) to identify group mobility patterns and provide steadier cluster formations.

ACKNOWLEDGMENTS

The research work presented herein has been co-funded by 75% from EU and 25% from the Greek government under the framework of the Education and Initial Vocational Training II, Programme Archimedes.